# FEASIBILITY OF DIRECT DISPOSAL OF SALT WASTE FROM ELECTOCHEMICAL PROCESSING OF SPENT NUCLEAR FUEL

Rob P Rechard, Teklu Hadgu, Yifeng Wang, Larry C. Sanchez,
*Sandia National Laboratories, Albuquerque, NM 87185-0747, USA*

Patrick McDaniel, Corey Skinner, Nima Fathi
*University of New Mexico, Albuquerque, NM, USA*

Steven Frank, Michael Patterson
*Idaho National Laboratory, Idaho Falls, ID, USA*

*The US Department of Energy decided in 2000 to treat its sodium-bonded spent nuclear fuel, produced for experiments on breeder reactors, with an electrochemical process. The metallic waste produced is to be cast into ingots and the salt waste further processed to form a ceramic waste form for disposal in a mined repository. However, alternative disposal pathways for metallic and salt waste streams are being investigated that may reduce the processing complexity. As summarized here, performance assessments analyzing the direct disposal the salt waste demonstrate that both mined repositories in salt and deep boreholes in basement crystalline rock can easily accommodate the salt waste. Also summarized here is an analysis of the feasibility of transporting the salt waste in a proposed vessel. The vessel is viable for transport to and disposal in a generic mined repository in salt or deep borehole but a portion of the salt waste would need to be diluted for disposal in the Waste Isolation Pilot Plant. The generally positive results continue to demonstrate the feasibility of direct disposal of salt waste after electrochemical processing of spent nuclear fuel.*

## I. INTRODUCTION

For experiments on reactors breeding plutonium, the US Department of Energy (DOE) tested a fuel with a layer of metallic sodium (Na). Directly disposing of this Na-bonded spent nuclear fuel (SNF) from these reactors without treatment in a geologic repository is not possible because of the potentially energetic reaction of the sodium metal with water to produce hydrogen gas and sodium hydroxide. Hence, DOE decided in 2000 to treat this Na-bonded SNF with an electrochemical process. The electrochemical process produces a metallic waste, which is mostly from the cladding, and a salt waste, which contains many of the actinides and fission products. [1] The current plan is for the salt waste to be further processed to form a ceramic waste form for disposal in a mined repository.

Yet other disposal paths are viable for the salt waste. Summarized here are performance assessments analyzing the direct disposal of salt waste (without treating it to form a ceramic waste form) in mined repositories in salt and deep boreholes in basement crystalline rock. This paper also discusses the feasibility of transporting a proposed disposal vessel and whether any issues would suggest that a smaller or larger size is more appropriate. This practical question addresses the issue as to whether it is necessary to develop plans and secure funding to modify facilities at Idaho National Laboratory (INL) in order for direct disposal of the salt waste to be feasible.

## II. BACKGROUND ON ER SALT WASTE AND DIRECT DISPOSAL OPTION

### II.A. Na-Bonded SNF

The Na-bonded driver fuel in the core of the experimental fast-spectrum breeder reactor consisted of highly enriched uranium (HEU) metal alloyed with 10 wt.% zirconium. The blanket fuel surrounding the reactor core for breeding the plutonium consisted of depleted uranium. Both the HEU metal-zirconium alloy driver and blanket fuel is surrounded by a layer of metallic Na and a layer of cladding to improve heat transfer. The cladding for the driver fuel was either D9, HT9, or 316 stainless steel. The cladding for the blanket fuel was predominately 304 stainless steel.

DOE has ~3.3 metric tons of heavy metal (MTHM) of driver SNF (~65% $^{235}$U enriched at discharge) and 22.4 MTHM of blanket SNF from the experimental breeder reactor (EBR-II) at INL,[1] and ~0.25 MTHM driver SNF from the Fast Flux Test Reactor (FFTF) at Hanford.[2]

### II.B. ER Salt Waste

For treating the experimental Na-bonded SNF, DOE decided to use two electrorefiners (ER) located in the Fuel Conditioning Facility (FCF) at INL. The Mark-IV ER would process the driver SNF. The Mark-V ER would process the blanket SNF.

In the ERs, a batch of chopped SNF is placed in anode metal baskets and immersed in a 500 $^{o}$C molten LiCl and KCl salt near its eutectic concentration. When current is passed through the metal baskets, fission products and actinides are oxidized and readily dissolve into the molten eutectic salt as chloride salts. The uranium is reduced to its metallic form and accumulates on the cathode. The uranium (~65% enriched) can be recovered, diluted to <20% enrichment by adding depleted uranium, and cast into ingots.

The irradiated cladding and most of the zirconium in the driver U-Zr SNF does not oxidize and dissolve into the salt, but rather remains in the anode basket.

Some noble metal activation products also remain with the anode basket such as Mo, Tc, Ru, Rh, and Pd. The existing metallic waste form (MWF) has been cast into 3 circular ingots in the furnace operating in the Hot Fuels Examination Facility (HFEF) at INL since 2012. Current plans call for the circular ingots to be placed in standard high-level waste (HLW) canisters (nominally 61 cm diameter and 3 m tall) for disposal (~5850 kg or 488 ingots from ~26 MTHM of EBR-II SNF.

Many of the fission yield products and actinides from the SNF, remain in the 400 kg of molten LiCl-KCl salt an ER. Eventually, the salt will be removed from each ER. Reasons include (a) reducing radionuclide concentration to avoid criticality concerns from $^{239}$Pu and possibly some $^{235}$U, (b) reducing the Na content to keep the melting point of the salt mixture below 500 °C, (c) reducing the actinides and fission products in the salt such that decay heat does not prevent solidification once removed from the ER, and (d) the ER is to be decommissioned.

## II.C. Baseline Pathway for Treating ER Waste

Although electrochemical processing was developed for treating the Na-bonded EBR-II SNF,[1] its use is far more general and could have great promise for treating other DOE-managed SNF, especially small amounts of HEU SNF.[3] Yet, for this technology to succeed in the US, a path to safe disposition of the resulting salt waste must be shown.

Because of the chloride salts, vitrification of the ER salt waste as borosilicate glass is not feasible. The current disposal pathway for the ER salt waste is to form a glass-bonded sodalite composite ceramic (referred to as a ceramic waste form or CWF).[4] The processing consists of removing the molten salt, solidifying, crushing, adding ground zeolite, and heating to 500 °C. This salt loaded zeolite is then mixed with a glass binder and heated to 925 °C in a furnace to form the CWF cylinders (Fig. 1).[3] The waste treatment equipment is to be located in the HFEF. Both the FCF and HFEF are at the Materials and Fuels Complex (MFC). As currently planned, 2 CWF cylinders are to be placed into a standard HLW canister and the HLW canister shipped to mined repository with other DOE-managed HLW and SNF. About 64 HLW canisters would be produced from processing EBR-II SNF.[2]

## II.D. Direct Disposal Pathways for ER Salt Waste

Examining other disposal paths for the ER salt waste is prudent because of (1) the de facto stoppage of the nation's first proposed mined repository in volcanic tuff at Yucca Mountain, Nevada, (2) the complexity of creating the ceramic waste form, (3) the large increase in mass volume of ER salt waste caused by CWF treatment, and (4) the limited space in the HFEF hot cells to accommodate the necessary equipment.

To elaborate upon the latter two points, 1.72 metric tons (MT) of EBR-II waste produces 50.95 MT of CWF.[4] Furthermore, the CWF equipment is large and would occupy a significant portion of the HFEF hot cell, which would likely preclude using HFEF for most other experimental purposes for the 3 to 5 years necessary to complete treatment.[4]

### II.D.1. Three Direct Disposal Pathways

The direct disposal option involves sending the ER salt waste directly to a repository without further treatment. Three direct disposal pathways exist for ER salt waste (Fig. 1). One pathway is to send the waste to a future deep borehole repository.

A second pathway is to send ER salt waste to a mined repository for commercial and/or defense waste. Until the proposed repository at Yucca Mountain, Nevada, came to a *de facto* stop in 2010 because of a lack of funding, this was the anticipated pathway, but with treatment of the ER salt waste to produce CWF. This pathway depends upon renewed Administrative and Congressional support for a commercial repository or a repository for only defense related waste.

### II.D.2. Disposal at WIPP

The third pathway is to send the portion of ER salt waste that meets the definition of defense related remote handled transuranic (RH-TRU) waste to the Waste Isolation Pilot Plant (WIPP) located in bedded salt in southern New Mexico (Fig. 1).

The *Waste Isolation Pilot Plant Land Withdrawal Act* sets several limits on TRU waste: (1) volume is limited to $1.76 \times 10^5$ m$^3$ ($6.2 \times 10^6$ ft$^3$); (2) total activity of RH-TRU is limited to $5.1 \times 10^6$ Ci; and (3) activity concentration for RH-TRU waste is limited to $2.3 \times 10^4$ Ci/m$^3$ (averaged over the canister volume).

## II.E. Vessels for ER Salt Waste

### II.E.1. Primary Salt Container

Direct disposal would involve placing the contaminated molten ER salt waste into a primary salt container (PSC) to cool. In concept, 3 sizes were considered for the PSC, but not all 3 sizes would make sense for the 3 disposal pathways. Specifically, the 3 PSC considered were (1) proposed baseline PSC for use in standard truck cask for shipment of RH-TRU to WIPP, (2) a smaller PSC for use in shipping contact handled transuranic (CH-TRU) waste to WIPP, and (3) a larger PSC for shipping via rail to a future mined repository for commercial and DOE-managed SNF.

### I.E.2. Proposed Primary Salt Container

A vessel for direct disposal of ER salt waste has been proposed, designed, and a prototype manufactured based on the current configuration of the HFEF and commonly used handling containers at the facility. As proposed, the PSC would be constructed of 316 stainless steel with a salt capacity of 43 kg.

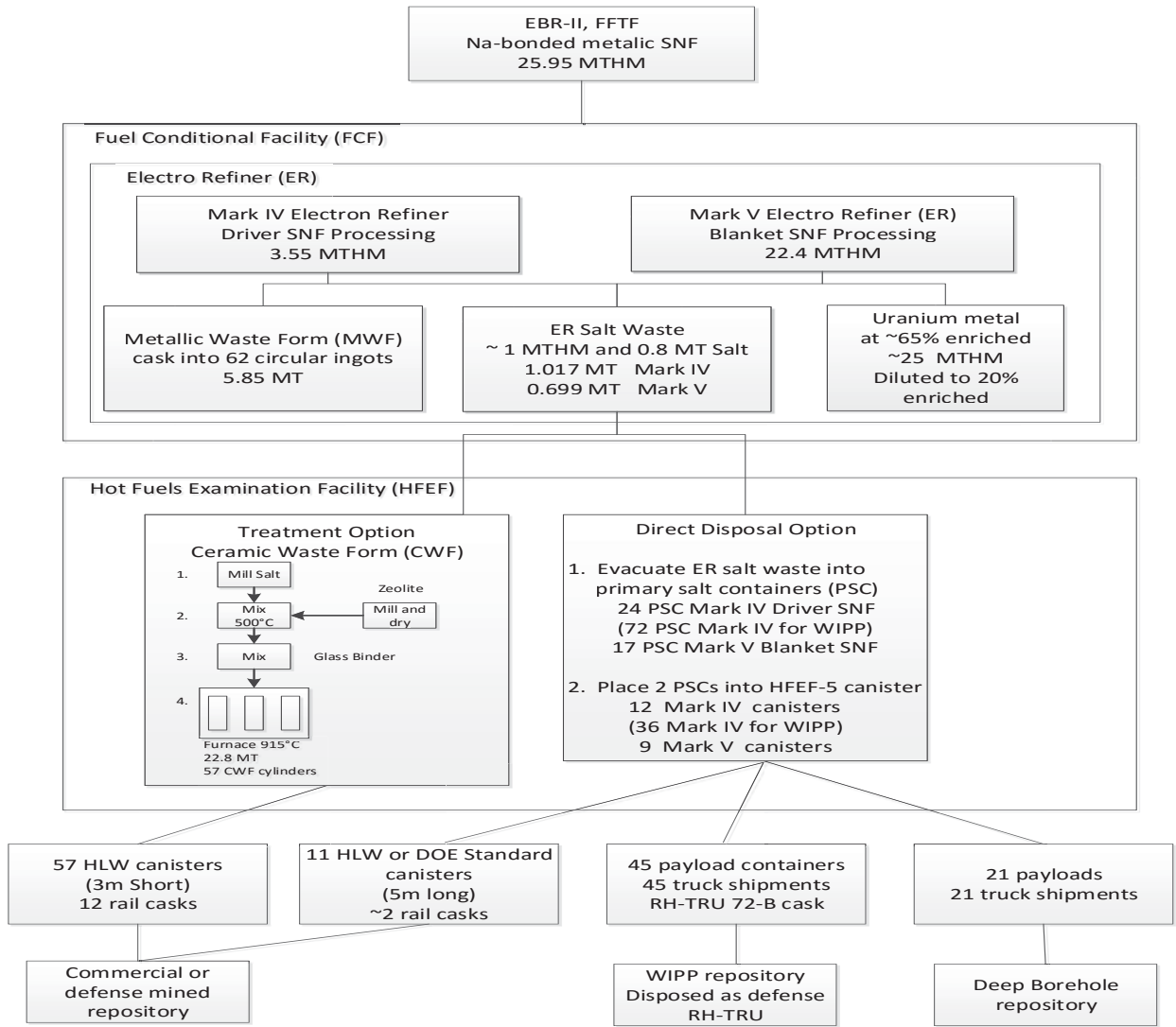

**Fig. 1. Various Pathways for Disposal of Na-Bonded SNF from EBR-II and FFTF.**

The estimated total mass of salt waste produced from the treatment of the 22.4 MTHM blanket SNF in the Mark-V ER is 699 kg.[4] The estimated total salt mass from processing the ~3.4 MTHM of the driver SNF in the Mark IV ER is 1017 kg. Hence, 17 PSCs are needed for the Mark-V ER salt waste and 24 PSCs are needed for the Mark-IV ER salt waste.

However, because of the statutory limit on activity concentration at WIPP ($2.3 \times 10^4$ Ci/m$^3$), Mark IV ER salt waste concentration would be constrained. The precise amount would depend on the packaging since the concentration is averaged over the volume of the disposal canister. As described later, the payload canister for the RH-TRU 72-B truck cask for WIPP is 0.9 m$^3$. Consequently, a PSC for the Mark IV must be a factor of 3 smaller or the salt waste diluted by a factor of 3, based on the activity of Mark IV waste (Table 1). In the latter situation, 72 PSCs would be produced for a total of 89 PSCs (Fig. 1).

**Table 1. Activity and thermal power in 2013 for 86 kg of ER salt waste in 2 primary salt containers.**

| Duration beyond 2013 | Mark IV | | | Mark V | | |
|---|---|---|---|---|---|---|
| | Activity (kCi) | Dilute Factor of 3 | Power (W) | Dilute Factor of 3 | Activity (kCi) | Power (W) |
| 0 | 160.6 | 53.5 | 327.2 | 109.1 | 48.6 | 24.9 |
| 1 | 119.6 | 39.9 | 322.0 | 107.3 | 4.8 | 24.6 |
| 10 | 63.7 | 21.2 | 261.8 | 87.3 | 3.3 | 21.9 |
| 30 | 43.6 | 14.5 | 166.1 | 55.4 | 2.3 | 17.5 |
| 100 | 7.8 | 2.6 | 26.2 | 8.7 | 0.6 | 12.0 |

*I.E.2. Proposed Primary Salt Container*

A vessel for direct disposal of ER salt waste has been proposed, designed, and a prototype manufactured based on the current configuration of the HFEF and commonly used handling containers at the facility. As proposed, the PSC would be constructed of 316 stainless steel with a salt capacity of 43 kg.

### II.E.4. Proposed HFEF Handling Canister

The inner container would then be placed in an HFEF-5 canister, a standard container used to move material within the HFEF. Hence, the proposed handling vessel consists of 3 nested cylinders around the waste. The HFEF-5 is constructed of 304 stainless steel with 32.38 cm diameter and 186.69 cm length.

### II.E.5. Alternative PSC for TRUPAC-II

A second option considered was to reduce the size of a PSC such that it could be used with the TRUPACT-II transportation cask, which was developed for shipping CH-TRU to WIPP. The PSC would be sized to fit into an 55-gallon drum using an overpack, such as the S200 pipe overpack.[5, p. 50]

### II.E.6. HLW or DOE Standard Canister Alternative

An alternative handling canister would be to place the HFEF-5 canister inside either a standard HLW canister or standard DOE canister for DOE-managed SNF. DOE will be developing many different types of baskets to support the DOE-managed SNF and a basket for the HFEF-5 canister could also be developed. A specific transportation cask has not been designated but would be chosen in conjunction with the shipment of DOE-managed SNF to a repository.

## III. SALT WASTE DISPOSAL PERFORMANCE

### III.A. Waste form and package performance

The role of the engineered barrier system (EBS) differs with the host geologic media. For all repositories, the container of the EBS provides important short-term radionuclide confinement for operations and when retrieval might be necessary. Long-term container and waste form performance is less important for a mined repository in salt (and clay/shale), because the geologic natural barrier system provides substantial long-term isolation. For salt repositories, the vessel used inside the transportation cask is likely to be sufficient for disposal. Disposal of the ER salt waste in a salt repository is ideal because the ER salt waste form is stable in the saturated brines of a salt repository.

Long-term radionuclide isolation by the EBS is more important for a mined repository in crystalline rock and volcanic tuff (when using a dose standard). Provided a sufficiently robust waste package is used, direct disposal of ER salt waste could also be viable for mined repositories in crystalline and volcanic rock, but would need to be verified with performance assessment (PA()) analysis

### III.B. Summary of Past PA Studies

Evaluation of the direct disposal of ER salt in a salt repository was initiated in 2010. In 2011 and 2012, work focused on laboratory studies of salt waste dissolution behavior in simulated salt repository brines.[3] In 2013, direct disposal of ER salt waste without further treatment was demonstrated as feasible for a generic salt repository without other waste.[6] In 2014, the PA included a detailed thermal analysis of waste package size and spacing, and disposal of ER salt waste along with commercial and DOE-managed SNF and HLW. Also, in 2014, a criticality analysis related to ER salt transportation and disposal showed no concerns.[6] Because of these studies, the direct disposal option was rated as promising in a 2014 DOE complex-wide evaluation of DOE-managed SNF and HLW.[2, Table 5-7] In 2015, the analysis shifted to the feasibility of deep borehole disposal and corresponding criticality analysis.

### III.C. Influence of Salt Waste Heat on WIPP

As currently envisioned, 1 or 2 HFEF-5 handling canisters would be in a standard payload canister. At WIPP, the standard payload canister would be put directly in the floor (or possibly the wall) of one of the disposal rooms. Previous analysis of the influence on performance of the ER salt waste has focused on the disposal in a mined repository in salt for commercial and/or defense-only waste or a deep borehole repository in crystalline rock for defense-only waste. For the mined repository, ER salt waste is relatively cooler than other waste (Fig. 2). However, at WIPP, the ER salt waste is relatively warmer than the CH-TRU. Nonetheless, the performance should be acceptable since ER salt waste would represent such a small portion of the total waste at WIPP.

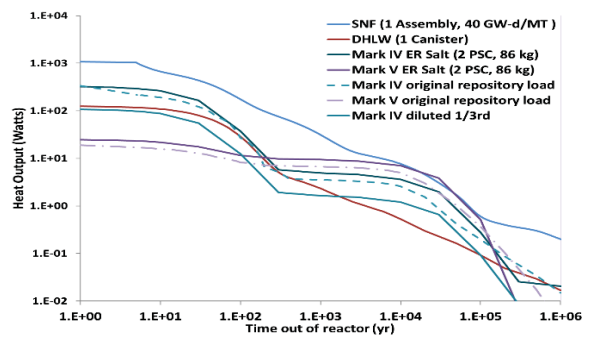

**Fig. 2. Heat output from ER salt waste compared to a HLW canister and an SNF assembly.**

To elaborate, the total heat load at WIPP was initially projected at ~136 kW in 1998.[7] Mark IV waste in 2033 would represent a total 3 kW (or a change of 2% from 136 kW) (Table 1), which will be difficult to observe in a PA. Granted, the additional heat load may be concentrated in the floor of only a few rooms; but this isolated and small concentrated heat load is still unlikely to have any adverse effects. In fact, during the certification of WIPP, a generic situation with a concentrated RH-TRU heat source was examined and found to be insignificant.[8] Nonetheless, a PA analysis of this particular situation will be necessary, if WIPP is chosen for disposing ER salt waste.

### III.D. Deep Borehole Disposal

The deep borehole disposal concept consists of drilling deep boreholes ~5 km into crystalline basement rock for permanent disposal of high level radioactive waste in the bottom 1 to 2 km. The disposal concept has been previously described.[6; 9] Advantages of the concept are that migration of radionuclides from the deep borehole would be severely restricted by the low permeability in deep crystalline rocks, limited interaction of deep fluids with shallower groundwater, and geochemically reducing conditions at depth, which limit the solubility and enhance the sorption of many radionuclides. The added advantage of direct disposal of ER salt waste is that the high salinity content of the waste would further restrict radionuclide mobility towards the accessible environment. This attribute is also applicable to disposal in other geological media.

*III.D.1. Thermal-Hydrology Analysis Model*

Thermal-hydrology analysis investigated density driven flow when emplacing ER salt in a deep borehole. To be consistent with past analysis, the original decay heat in a RH-TRU 72-B payload disposal package (86 kg) for the ER salt waste were used in the simulations described here (Fig. 2).[6]

The analysis includes detailed three-dimensional modeling using the numerical code PFLOTRAN to assess magnitude and direction of fluid movement in the vicinity of the borehole because of density differences between the highly saline ER salt and the surroundings.[10] To accentuate differences, the surrounding crystalline rock had salinity <990 kg/m$^3$. The ER salt canisters were placed over 34 m in a 200-m section at the top of the disposal zone at a depth of 2900 m (bottom at 3100 m). Sedimentary overburden would normally be present, but the model domain is exclusively in the crystalline rock.

*III.D.2. Thermal Analysis Results for Deep Borehole*

PFLOTRAN simulations were conducted for a total simulation time of 10$^6$ years. At 3000 m, there is a small thermal perturbation at early time as a result of the decay heat of the ER salt waste. As decay heat decreases, the temperature approaches the *in situ* temperature associated with geothermal gradient. (Fig. 3). The vertical groundwater flow is minimal, particularly when compared to perturbations caused by emplacement of SNF in previous calculations.[6]

The concentration profile at 3000-m depth indicates movement of fluid by advection. A small peak is reached at around 250 years and then much larger concentrations at later times (Fig. 3).

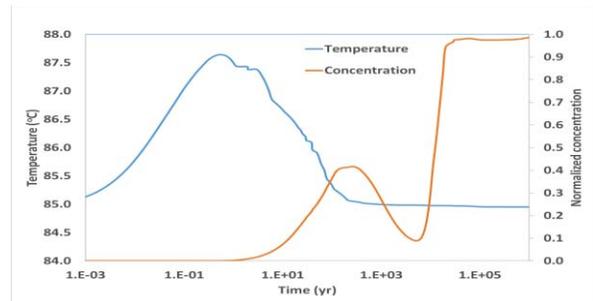

**Fig. 3. Temperature and salt concentration as function of time in borehole at 3000-m depth.**

The analysis results suggest that any contaminant plume released from the ER salt waste emplaced in a segment of borehole would move downward due to the density flow (**Error! Reference source not found.**). By 50,000 years, the plume has traveled ~100 m downward.

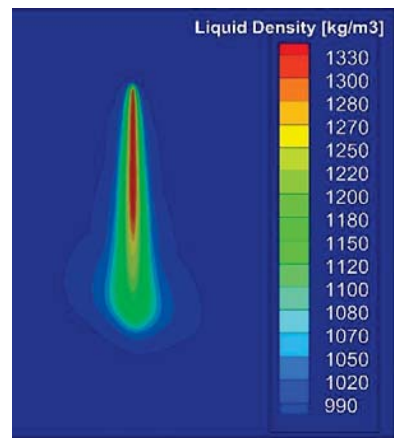

**Fig. 4. Liquid density at 20,000 years.**

## IV. TRANSPORTATION OF ER SALT WASTE

Three issues are addressed related to currently proposed containers and available transportation casks (1) shielding necessary to reduce doses to acceptable levels; (2) the criticality potential and the ease which it can be shown to be inconsequential when amending a transportation cask certificate of compliance (CoC), and (3) temperatures of the containers in relation to acceptable cask limits. Usually, the dose and thermal limits are the most important in determining the practical feasibility of shipping the ER salt waste.

### IV.B. Transportation Casks for Direct Disposal

*IV.B.1. RH-TRU 72-B Cask for Truck Shipments to WIPP and Borehole Repository*

A proposed truck cask for shipping the HFEF-5 handling canister to WIPP or the deep borehole repository is the existing RH-TRU 72-B transportation cask, (NRC Certification 9212), which was designed to transport RH-TRU waste.[11] The cask can transport a payload (including the payload canister) weighing

3628 kg. The inner payload canister is composed of either carbon or 304 stainless steel with 66 cm outer diameter and 306 cm length. The internal volume is ~0.9 m$^3$ (Fig. 5).[12]

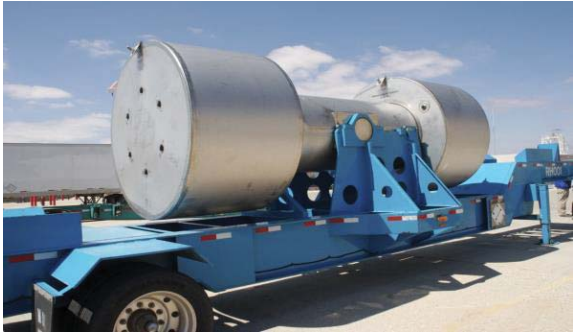

**Fig. 5. RH-TRU 72-B cask for remote-handled TRU.**

Calculations were performed with one and two HFEF-5 cans placed within the RH-TRU 72-B cask, and centrally suspended in dunnage within the payload vessel. Several options for dunnage were considered: (1) redwood cellulose (which can reasonably represent Celotex), (2) sand, (3) salt, and (4) a commercial product, polysiloxane with bismuth. Dunnage is assumed to be placed outside the HFEF-5 canister and within the lead shielding circumference of the payload canister. For the cases with two HFEF-5 canisters, the lead or depleted uranium shielding shell was treated as a circular envelope surrounding the containers. The dose calculations were performed including the optional steel shield in the HFEF-5 canister.

*IV.B.2. TRUPAC-II Cask for Truck Shipments to WIPP and Borehole Repository*

The TRUPAC-II is an alternative transportation cask currently available for shipping waste to WIPP. The TRUPACT-II transportation cask is not designed to provide significant gamma or neutron shielding. Hence, the payload contents must be shielded. The approved payloads are the thin-walled 55-gallon drum, 85-gallon drum, 100-gallon drum, standard waste box, and 10-drum overpack. The 55-gallon drum is the basis of several payloads with additional internal components to provide either criticality control or additional shielding such as the S200 pipe overpack. Each TRUPACT-II cask can transport fourteen 55-gallon drums. A truck can carry 3 TRUPACT-II casks.

*IV.B.3. Cask for Transport to Mined Repository for Commercial or Defense SNF*

In general, the thermal history for 2 PSC of Mark IV ER waste is like that of a generic HLW canister (Fig. 2). Hence, casks designed to ship vitrified HLW can ship ER waste. The newer casks designs are for rail transport of large quantities of SNF and HLW (e.g., 5 canisters of HLW), a weight far greater than in 5 packages of ER waste containing 2 PSC per package. The maximum heat load is ~20 kW, which roughly corresponds to 65 packages of ER waste (Table 1).

**IV.C. Shielding Necessary in Payload Canister for Transport to WIPP and Borehole Repository**

*IV.C.1. NRC Dose Limitations for Casks*

For normal conditions of transport, the Nuclear Regulatory Commission (NRC), under 10 CFR 71, requires that (1) dose is < 2 mSv/h (200 mrem/h) on surface of the transportation cask and (2) dose is < 0.1 mSv/h (10 mrem/hr) 2 m from cask surface (§71.47).

*IV.C.2. Calculation Method*

The required thickness of lead or depleted uranium shielding in addition to shielding provided by the dunnage and nested steel containers was determined by trial and error to find acceptable doses at the exterior of the transportation cask. The dose calculations were performed with the MAVRIC-MONACO code sequence of SCALE v6.1 The source term was Mark IV (without a factor of 3 dilution).

*IV.C.3. Doses Outside RH-TRU 72-B*

When shipping one HFEF-5 canister with the optional pipe shielding in the RH-TRU 72-B transportation cask with redwood dunnage (Fig. 6), ~ 0.5 cm layer of addition lead shielding or 0.35 cm layer of depleted uranium shielding is required. When salt, sand, or commercially available polysiloxane with bismuth is used as a dunnage, no addition lead or depleted uranium shielding is necessary.

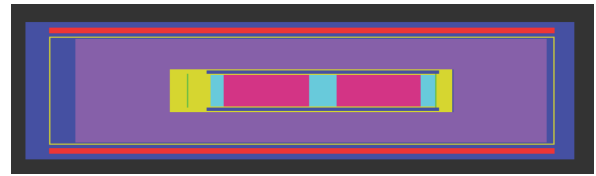

**Fig. 6. RH-TRU 72-B cask with dunnage (purple region) and additional shielding when transporting one HFEF-5**

When shipping two HFEF-5 canisters in the RH-TRU 72-B transportation cask with redwood dunnage, ~1 cm thick layer of addition lead shielding or ~0.65 cm thick layer of depleted uranium shielding is required. When salt is used as a dunnage, an additional shielding layer of lead ~0.75 cm thick or depleted uranium 0.33 cm thick is necessary. With polysiloxane and bismuth as dunnage, no addition lead or uranium shielding is necessary.

*IV.C.4. Doses Outside TRUPACT-II Cask*

An overpack for a small PSC that is adequate for transporting ER salt waste in 55-gallon drums used in TRUPACT-II cask does not currently exist. Adding lead shielding to an existing two variations of the S200 pipe overpack would leave space for only 4.4 or 7.4 kg of ER waste and the mass of lead would exceed the currently approved limits. Use of the TRUPACT-II cask for transporting ER salt waste will require (1) developing a specific overpack for a small alternative PSC, (2) testing the overpack in a drop test and fire test (possibly numerically), and (3) modifying the CH-TRAMPAC (documentation associated with the safety analysis report—SAR—for TRUPACT-II)

**IV.D. Lack of Criticality Potential**

In general, no known characteristic of ER waste would make transportation or disposal very difficult. However, the Pu concentration in ER salt waste is higher than commercial SNF and defense related high level waste (i.e., ~3 wt.% $PuCl_3$ in the Mark-IV ER salt and ~6 wt.% $PuCl_3$ in the Mark-V ER salt), which would require separate analysis (1) as to the potential for criticality during storage, transportation, and disposal, and (2) any safeguards and security risks.

*IV.D.1. Calculation Method*

The potential of criticality in the ER salt waste while transported in an RH-TRU 72-B cask was estimated with the XSDRNPM module of SCALE. The multiplication factor for an infinite medium of ER salt waste ($k_{inf}$) was <0.7. An effective multiplication factor for ER waste in the actual configuration ($k_{eff}$) was also calculated using the KENO-IV Monte Carlo capability in SCALE.

*IV.D.2. Low Criticality Potential*

The RH-TRU 72-B CoC limits Pu fissile mass to 315 fissile gram equivalent (FGE) or 370 FGE if ≥25 g of $^{240}$Pu is present.[11; 13] However, $k_{eff}$<<1, regardless of the dunnage and shielding used, because the fissile material is uniformly dispersed in a neutron absorbing chloride salt in the ER salt waste. Hence, criticality during transportation is not of practical concern, but will require an CoC amendment.

*IV.D.3. Physical Protection*

Safeguards and security may have to be addressed because of the 2 to 5 kg of $^{239}$Pu in 2 PSCs. However, the attractiveness of the ER salt waste to theft, diversion, or sabotage is very low.

**IV.E. Influence of Shielding on RH-TRU 72-B Temperatures**

*IV.E.1. Heat Load Limits*

The WIPP waste acceptance criteria (WAC) limit all payload canisters to that specified in the CH-TRAMPAC and RH-TRAMPAC. The maximum thermal heat load for shipping non-combustible metallic waste used in the SAR for the RH-TRU 72-B cask is 300 W. The maximum thermal heat load for a one drum container for the TRUPACT-II cask is 50 W. This limit is in place because some canisters shipping to WIPP may contain combustible waste, and it is desirable to limit the potential for radiolysis generating hydrogen gas while being transported.

*IV.E.2. Thermal Load of Source*

As noted previously, the thermal history for 2 PSC of Mark IV ER waste is similar to generic HLW or relatively cool DOE-managed waste (Fig. 2). Granted, the projected initial thermal power output is 327.2 W in 2 PSC in 2013 (Table 1). But after a little more than 4 years (2017), the thermal output for 2 PSC in 1 HFEF-5 canister will be less than the 300-W limit for non-combustible, metallic waste in the RH-TRU 72-B cask.

After 10 years, up to 16 kg of the Mark IV salt waste could be placed in a small PSC container and meet the 50 W limit for TRUPACT-II (Table 1).

*IV.E.3 Influence of shielding on cask temperatures*

Based on a finite-element thermal model of the RH-TRU 72-B the thermal loads from the proposed ER salt waste inside HFEF-5 canister with additional shielding is acceptable for a 300-W source under normal conditions of transport (NCT). For example, the temperature on the cask surface is ~34 ºC above the ambient temperature of 37.8 ºC (Fig. 7).

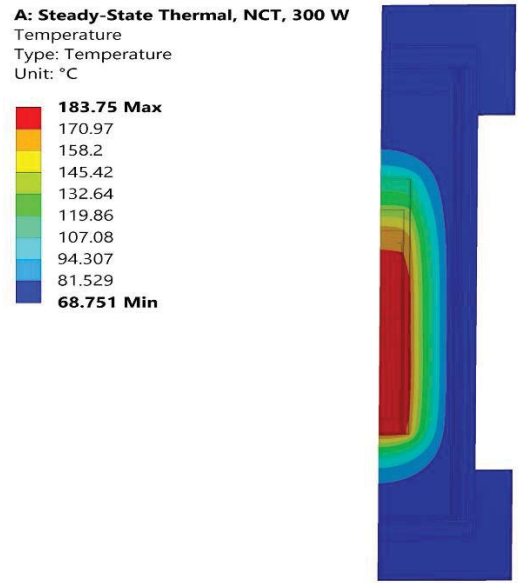

**Fig. 7. Temperatures within RH-TRU 72-B with 2 PSCs at 300 W inside a standard payload canister as calculated by ANSYS (Rev 17.1) for normal transport conditions.**

**V. CONCLUDING REMARKS**

As summarized here, PAs analyzing the disposal of ER salt waste directly (without treating it to form a glass ceramic) show that both mined repositories in salt and deep boreholes in basement crystalline rock can easily accommodate the ER salt waste. For deep borehole disposal, a container diameter must be small; but in general, these PAs verify the usual adage that if

SNF or high-level radioactive waste (HLW) can be transported to the repository under US regulations, it can be disposed under US regulations, provided social-political limitations on the type and amount of waste are met.

Work in 2016 began developing the basis to show the operational feasibility of the direct disposal of ER salt waste. Here, the emphasis has been on the transportability of the proposed vessels. Calculations of dose, criticality, and temperatures showed that the proposed option was feasible without the need to secure funding to modify the facility.

As regards WIPP, two PCS would be loaded and sealed into HFEF-5 canisters. In turn, one or two HFEF-5 canisters could be loaded into a RH-TRU 72-B transportation cask for shipment. Although the ER salt waste is far below any criticality concern, funding may be required sometime in the future to amend the RH-TRAMPAC and indirectly the WIPP WAC to allow higher fissile amounts in a container.

The RH-TRAMPAC and indirectly the WIPP WAC would also need to be modified to allow ~90 W (Table 1) heat loads for 45 payload canisters containing the Mark IV ER salt waste for disposal in 2023 (Fig. 1). Furthermore, WIPP specific PA codes will need to be used to provide the basis that ER salt waste does not materially alter the safety case.

Smaller and larger PCS were also considered to determine the feasibility and if any advantages occurred. One path forward is to develop a small PSC that will hold between 4 and 8 kg of ER salt waste. Between 400 and 200 PSCs would be produced. The primary advantage of using a smaller PSC would be to avoid having to dilute the Mark IV ER salt waste to shipment and disposal in the RH-TRU 72-B payload canister and make use of the readily available TRUPACT-II transportation cask.

Existing overpack containers for the standard 55-gallon drum payload have insufficient shielding for ER salt waste. This option would require developing an inner container for the 55-gallon drum payload and amending the TRUPACT-II CoC. Although feasible, this option will have to be compared with the alternative option of modifying the RH-TRU 72-B CoC to accept the larger fissile masses in the currently proposed PSC.

The primary advantage of a larger PSC would be to make efficient use of canisters for DOE-managed SNF, which would be shipped to a repository along with the entire inventory of DOE-managed SNF. The disadvantage to this approach is that the disposition of ER salt waste would be linked to that of the DOE-managed SNF. Furthermore, this approach would reduce the ability to use the other two disposal pathways: WIPP or deep boreholes (i.e., the current proposed size for the PSC offers sufficient flexibility for a viable disposal pathway).

## ACKNOWLEDGEMENTS

Sandia National Laboratories (SNL) is a multi-mission laboratory operated by Sandia Corporation, a wholly owned subsidiary of Lockheed Martin Corporation, for the US Department of Energy (DOE) National Nuclear Security Administration under contract DE-AC04-94AL85000. The statements in this paper are those of the authors and do not necessarily reflect the policies of DOE, SNL, or INL.